# Magnetic Phase Diagram of $Mn_{3+x}Sn_{1-x}$ Epitaxial Thin Films: Extending the Anomalous Hall Effect to Low Temperatures via Intrinsic Alloying

K. Gas[1,2,a)], J.-Y. Yoon[3,4], Y. Sato[3,4], H. Kubota[3,4], P. Dłużewski[2], S. Kret[2], J. Z. Domagala[2], Y. K. Edathumkandy[2], Y. Takeuchi[3,5,6], S. Kanai[1,3,4,6,7,8,9], H. Ohno[1,3,7,10], M. Sawicki[2,3], and S. Fukami[1,3,4,7,10,11,a)]

[1] Center for Science and Innovation in Spintronics, Tohoku University, Katahira 2-1-1, Aoba-ku, Sendai 980-8577, Japan
[2] Institute of Physics, Polish Academy of Sciences, Aleja Lotnikow 32/46, PL-02668 Warsaw, Poland
[3] Laboratory for Nanoelectronics and Spintronics, Research Institute of Electrical Communication, Tohoku University, Katahira 2-1-1, Aoba-ku, Sendai 980-8577, Japan
[4] Graduate School of Engineering, Tohoku University, Katahira 2-1-1, Aoba-ku, Sendai 980-8577, Japan
[5] ICYS, National Institute for Materials Science, Tsukuba, Japan
[6] PRESTO, Japan Science and Technology Agency (JST), Kawaguchi, Japan
[7] Advanced Institute for Materials Research, Tohoku University, Katahira 2-1-1, Aoba-ku, Sendai 980-8577, Japan
[8] DEFS, Tohoku University, Katahira 2-1-1, Aoba-ku, Sendai 980-8577, Japan
[9] National Institutes for Quantum Science and Technology, Takasaki, Japan
[10] Center for Innovative Integrated Electronic Systems, Tohoku University, Katahira 2-1-1, Aoba-ku, Sendai 980-8572, Japan
[11] Inamori Research Institute for Science, Kyoto 600-8411, Japan

[a)]Authors to whom correspondence should be addressed: gas.katarzyna.a2@tohoku.ac.jp and s-fukami@riec.tohoku.ac.jp

**Abstract**

Antiferromagnets with broken time-reversal symmetry, such as $Mn_3Sn$, have emerged as promising platforms for exploring topological and correlated electron physics. $Mn_3Sn$ is known to show two magnetic phase transitions: a non-collinear inverse triangular antiferromagnetic (IT-AFM) spin configuration is formed below its Néel temperature ($T_N \cong 420$ K), whereas at $T_1$ that usually locates below room temperature, it transits to an incommensurate spin state. Accordingly, intriguing properties such as a strong anomalous Hall effect, observed from $T_N$ to $T_1$, disappear below $T_1$, limiting its utility at low temperatures. While bulk $Mn_3Sn$ has been extensively studied, the magnetic phase transitions and their tunability in thin films remain largely unexplored. Here, we investigate the magnetic and magneto-transport properties of $Mn_{3+x}Sn_{1-x}$ epitaxial thin films prepared by magnetron sputtering, systematically varying the Mn-Sn composition. Our results reveal that intrinsic alloying with Mn provides us with a handle to tune $T_1$, with the IT-AFM phase stabilized down to liquid helium temperatures for $x > 0.15$. From a magnetic phase diagram for epitaxial thin films, we also find a consistent magnetic anomaly ~55 K below $T_N$, accompanied by thermal hysteresis. Furthermore, the reduction of $T_N$ in thin films relative to bulk values is shown to correlate with lattice parameter changes. These findings extend the accessible temperature range for $Mn_3Sn$'s topological properties, paving the way for novel applications and further investigations into the interplay of spin, lattice, and electronic degrees of freedom in thin-film geometries.

## I. INTRODUCTION

Antiferromagnets with macroscopically broken time-reversal symmetry emerge as promising materials for spintronics due to their unique properties, which overcome some of the limitations of ferromagnetic materials.[1–6] $Mn_3Sn$ is a representative system in antiferromagnetic (AFM) spintronics. It crystallizes in a hexagonal $D0_{19}$ structure belonging to space group $P6_3/mmc$. Below its Néel temperature ($T_N \cong 420$ K) $Mn_3Sn$ adopts a non-collinear inverse triangular antiferromagnetic (IT-AFM) spin configuration within its kagome lattice, with spins forming $120^o$ angles and a negative vector chirality in the basal $a$-$b$ plane (0001).[7–9] The magnetic structure of $Mn_3Sn$ is characterized by the ferroic ordering of cluster magnetic octupoles, which serves as the key order parameter breaking time-reversal symmetry.[10,11] This symmetry breaking leads to the emergence of significant Berry curvature in momentum space, particularly near Weyl points.[12,13] Consequently, despite its negligible net magnetization, $Mn_3Sn$ displays substantial non-trivial transport phenomena, including large anomalous Hall effect (AHE),[1] the anomalous Nernst effect,[14] the second-order Hall effect[15] and the tunneling magnetoresistance effect,[16] all originating from is unique band structure, highlighting its potential for spintronic applications. Furthermore, the slight canting of Mn moments within the (0001) plane produces a small ferromagnetic moment (~0.002 $\mu_B$/Mn, $\mu_B$ is the Bohr magneton),[8,17] allowing external magnetic fields to control the antiferromagnetic order. This distinctive combination of magnetic and transport properties offers immense potential for the electrical manipulation of the AFM order, analogous to that observed in ferromagnets.[18–20]

For the investigation of the functionalities of non-collinear antiferromagnets, an in-depth understanding of their magnetic phase transitions is of high importance. In the case of $Mn_3Sn$, the Néel temperature ($T_N$) is a critical parameter, particularly in the context of Joule heating, as it defines the operational temperature range of the material.[21] Surface effects alone can reduce the already relatively low $T_N$ of $Mn_3Sn$ crystals by as much as 10 K,[22] highlighting the need to explore the factors influencing $T_N$ in different material configurations. Below room temperature, $Mn_3Sn$ undergoes a magnetic phase transition at $T_1$ (which can be as close to room temperature as 275 K) from an IT-AFM spin configuration to an incommensurate magnetic phase.[9,23] This transition is not only significant for understanding low-temperature physics but also for its impact on properties like AHE and weak magnetization, both of which are suppressed below $T_1$.[24] The $T_1$ transition is also attracting attention due to its origin based on charge and spin density-waves instabilities associated with Fermi-surface nesting[25] of flat bands.[23] This nesting phenomenon is highly sensitive to variations in the chemical potential and the bond lengths between Mn atoms, which jointly influence the band structure and may explain the sample-dependent incommensurability observed in $Mn_3Sn$.[1,9,24,26–28] Density functional theory calculations suggest that the Fermi energy $E_F$ is strongly influenced by the amount of excess Mn content, with $E_F$ increasing by approximately 6 meV for each percent of substitutional Mn atoms.[12] These effects provide valuable insights into the interplay between topology and electron correlations and even open pathways for exploring novel topological phases in magnetic materials. However, the suppression of the AHE below $T_1$ limits $Mn_3Sn$'s utility in investigating the low-temperature physics of octupolar order.[29]

While magnetic phase transitions like $T_N$ and $T_1$ have been reasonably well studied in bulk $Mn_3Sn$,[26] their behavior and tunability in thin-film structures are not well understood. Thin films often exhibit significant differences in transition temperatures compared to their bulk counterparts due to structural factors such as surface effects, epitaxial strain, and the presence of heterogeneous substrate and seed layers. Despite the challenges of experimental investigation – particularly the difficulty of detecting tiny magnetizations in AFM materials – studies of these phase transitions in thin films are essential for a complete understanding of their magnetic and electronic properties and an exploration of device functionalities.

It is therefore timely and important to systematically study the magnetic transition temperatures in thin films of $Mn_3Sn$ with varying Mn-Sn composition. This will facilitate the identification of the conditions that result in the formation of the magnetic phase showing intriguing device functionalities, i.e., the IT-AFM spin configuration. Additionally, it will enable the study of novel physics in strongly correlated systems that usually arise at lower temperatures. To this end, we investigate the temperature dependence of magnetic and magneto-transport properties of $Mn_{3+x}Sn_{1-x}$ thin films, prepared by intrinsic alloying understood as the incorporation of excess Mn atoms into the $Mn_3Sn$ lattice. This self-regulating adjustment of the Mn-Sn ratio inherently tailors the material's composition and properties without the need for external dopants, by Mn within the frame of the sputtering deposition technique. We establish precise values of the magnetic transition temperatures showing that for $x > 0.15$ the inverse triangular spin configuration which hosts the strong AHE can be stabilized down to the liquid helium temperature. Our results are summarized into the magnetic phase diagram for epitaxial thin films in which a magnetic anomaly below the paramagnetic-antiferromagnetic transition has been identified.

## II. SAMPLES AND METHODS

$Mn_{3+x}Sn_{1-x}$ layers are deposited on single crystal MgO(110) substrates by DC/RF magnetron sputtering according to previously elaborated protocols.[30] The sample stacks, depicted in Fig. 1(a) consist of, from the substrate side, W (2 nm)/Ta (3 nm) seed layer followed by 40 nm of $Mn_{3+x}Sn_{1-x}$ and terminated by MgO (1.3 nm)/Ru (1 nm) cap. W, Ta, $Mn_{3+x}Sn_{1-x}$ and Ru are deposited at 400 °C and MgO is deposited at room temperature. All steps are done under a base pressure of less than $1 \times 10^{-6}$ Pa. The established stack structure allows for obtaining $(10\bar{1}0)$-oriented ($m$-plane in hexagonal short-hand notation) epitaxial $Mn_{3+x}Sn_{1-x}$ films with the kagome plane perpendicular to the film plane.[30–32] The deviation $x$ from the stoichiometric composition in the $Mn_{3+x}Sn_{1-x}$ series is set by alternating 50 sec of Mn-Sn co-sputtering and $t_{Mn}$ (7.5 - 35.0 sec) of Mn single-target sputtering.[33,34] The consequent Mn-Sn composition is determined by inductively coupled plasma mass spectrometry. We adopt the $Mn_{3+x}Sn_{1-x}$ notation following Refs.[33,34]. The resulting magnitudes of $x$ show monotonical increase with $t_{Mn}$, as displayed in Fig. 1(b), and can be approximated by a linear function.

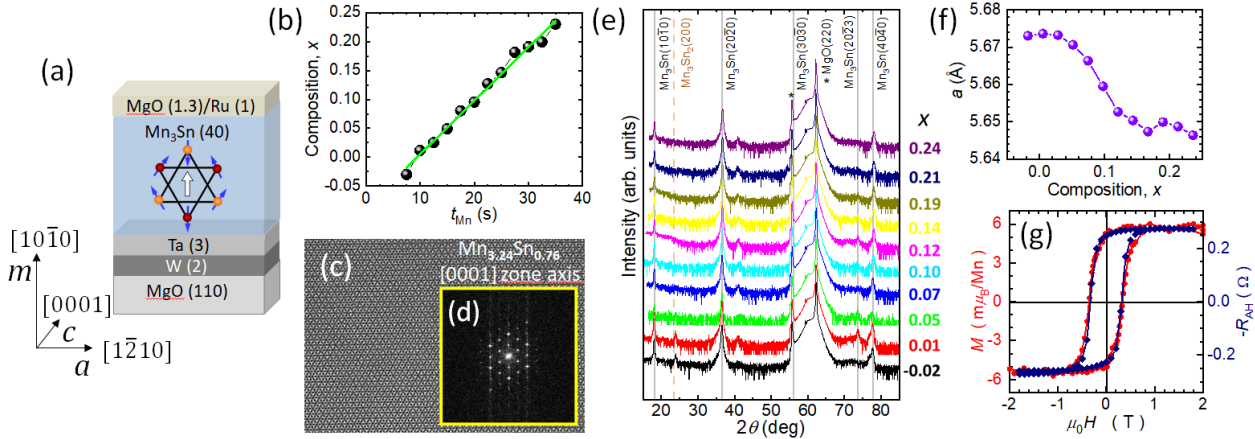

**FIG. 1**. Sample layout and structural properties of $Mn_{3+x}Sn_{1-x}$ films. (a) Schematic of the sample stack with thicknesses (in nm) of the constituent layers indicated in the parentheses. The inverse triangular spin arrangement in $Mn_3Sn$ is depicted in the central part. The red and orange spheres represent Mn atoms in two consecutive kagome planes. The blue arrows represent the magnetic moments of Mn atoms. The cluster magnetic octupole is represented by the white arrow. The coordinate system defines the crystallographic relations of $Mn_3Sn$. (b) (Black bullets) Exceeding Mn composition, $x$, as a function of Mn single target sputtering time $t_{Mn}$. The green line is the linear fit to the data. (c) Transmission electron microscope image in the [0001] zone axis of $Mn_{3.24}Sn_{0.76}$ film. (d) 2D diffraction pattern taken from a central fragment of (c). (e) Diffraction patterns in $2\theta/\omega$ of the samples. (f) $a$ lattice parameter dependence on $x$. (g) Comparison of magnetization, $M$ (red circles), and anomalous Hall resistance, $R_{AH}$ (navy diamonds), magnetic field $H$ loops ($H \parallel m$) for $Mn_{3.07}Sn_{0.93}$ film at $T = 300$ K. Note the inverse polarity of $R_{AH}(H)$ loop with respect to $M(H)$.

The homogeneous $D0_{19}$ structure of $Mn_3Sn$ is obtained through post-deposition annealing at vacuum by heating the samples to 600 °C at a rate of 10 °C/min, holding for 90 min at the target temperature, and subsequent cooling to room temperature. The material is typically divided into $5 \times 5$ or $10 \times 5$ mm$^2$ specimens. The former is used for magnetic and/or electrical measurements in van der Pauw configuration, whereas the latter is used for the electrical measurements with lithographically processed double-bridge-type Hall bar structures. Magnetic and electrical measurements in a wide temperature $T$ range from 5 to 400 K are performed in perpendicular orientation of magnetic field $H$ ($H \parallel m$) using Quantum Design's superconducting quantum interference device (SQUID) magnetometer MPMS XL5 and Physical Property Measuring Systems (PPMS), respectively. We strictly follow the experimental code of sensitive magnetometry.[35,36] DC electrical excitation is used for electrical studies. Titan Cubed 80-300 transmission electron microscope (TEM) operating with accelerating voltage 300 kV is used for the study. The crystal structure and lattice parameters are determined using an x-ray diffractometer with a wavelength 1.5406 Å of Cu K$_{\alpha1}$ line.

## III. RESULTS AND DISCUSSION
### A. Structural, magnetic and magneto-transport characterizations at room temperature

Cross-sectional TEM image shown in Fig. 1(c) and a corresponding electron diffraction pattern [Fig. 1(d)] for a stack with $x = 0.24$ corroborate the synthesis of high-quality epitaxial films of the *m*-plane $D0_{19}$-Mn$_{3+x}$Sn$_{1-x}$ even with high Mn content. X-ray diffraction patterns for all of the layers in the study are collected in Fig. 1(e). The positions of the $20\bar{2}0$ reflections allow to trace the evolution of the *a* lattice parameter with $x$ [shown in Fig. 1(f)]. Our findings confirm that *a* generally decreases with $x$, aligning with the dependency previously established for epitaxial thin films[33] and bulk polycrystalline samples.[26] The decrease of *a* implies the substitutional Mn incorporation into the Sn site, rather than assuming interstitial positions. However, the flattening of the $a(x)$ dependency seen above $x = 0.15$ indicates that not all Mn atoms provided during intrinsic alloying are substitutionally incorporated into the lattice. This suggests that a thermodynamic limit for Mn incorporation may have been reached.[26] It is supported by the observation of sub-micrometer-sized Mn-rich deposits on the surface of these films (shown in Fig. S1 of the Supplementary Information). For near stoichiometric layers additional diffraction peaks from $B8$-Mn$_3$Sn$_2$ are also observed, consistent with.[34]

We then measure the magnetization (*M*) and the anomalous Hall resistance ($R_{AH}$). A typical example for the Mn$_{3.07}$Sn$_{0.93}$ film at $T = 300$ K is shown in Fig. 1(g), where a strong magnetic contribution from MgO substrate[37] is removed. A good consistency in response to the applied magnetic field is confirmed. We highlight two challenges here. The first is that the diamagnetic signal from a typical $5 \times 5 \times 0.5$ cm$^3$ MgO substrate dwarfs that of the AFM layers in question, as shown in Fig. S2 in the Supplementary Information. Namely, above $10^{-4}$ emu and below $5 \times 10^{-6}$ emu at $\mu_0 H = 1$ T, respectively ($\mu_0$ is the permeability of vacuum). Moreover, the latter decreases by another order of magnitude as *T* approaches $T_N$. Second, the presence of a relatively strong soft ferromagnetic-like component in the MgO substrates alone "dissects" the square-like hysteresis loop of Mn$_3$Sn, leading to false double switching behavior and saturation levels. Overcoming these experimental hurdles proved to be essential in order to determine all the details of the magnetic phase diagram reported in this study. Our experimental approach is outlined in section S. II of the Supplementary Information (Refs.[38–41]). A clear relationship *M* and $R_{AH}$ further supports the general understanding that the applied magnetic field couples with a small, uncompensated weak ferromagnetic-like moment (WFM), which forms within the hexagonal crystal planes. On the other hand, results in Fig. 1(g) indicate that our $Mn_3Sn$ layers exhibit spontaneous magnetization magnitudes (about 2.5 mT or 20 m$\mu_B$ per formula unit at 300 K) approximately three times larger than those of bulk samples.[1,42] This enhanced magnetization is likely due to the tensile epitaxial strain, which was shown to significantly affect the magnitude of WFM.[43]

## B. Temperature dependent magnetization and magneto-transport measurements

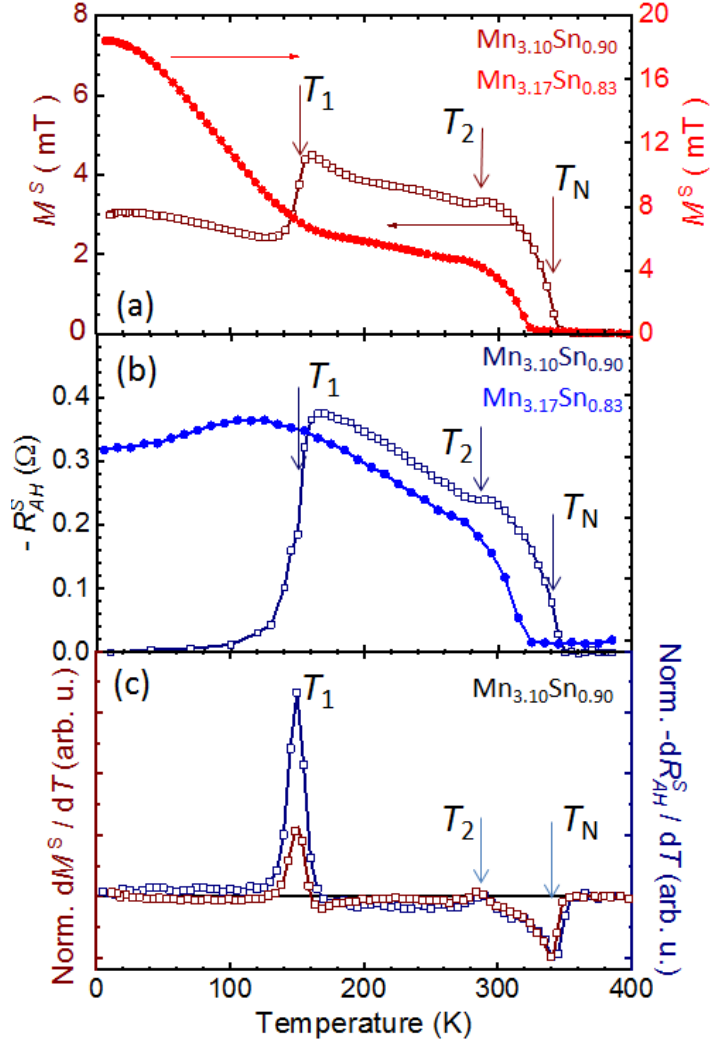

**FIG. 2.** (a,b) Temperature dependence of the spontaneous magnetization, $M^S(T)$ and anomalous Hall resistance, $R_{AH}^S(T)$, respectively, recorded on cooling in an external magnetic field $\mu_0 H_{FC} = 10$ mT for $Mn_{3.10}Sn_{0.90}$ (open squares) and $Mn_{3.17}Sn_{0.83}$ (bullets). (c) Temperature derivatives $dM^S(T)/dT$ and $dR_{AH}^S(T)/dT$ from (a) and (b) for $Mn_{3.10}Sn_{0.90}$. The positions of the Néel, $T_N$, and $T_1$ transitions, and an anomaly which we mark as $T_2$ are indicated by the arrows for the $Mn_{3.10}Sn_{0.90}$ only (for clarity). The $T_2$ anomaly is discussed in Section D.

The information extracted from temperature dependences of the spontaneous magnetization $M^S(T)$ and spontaneous anomalous Hall resistance $R_{AH}^S(T)$ provides us with an overview of the magnetic properties of $Mn_{3+x}Sn_{1-x}$ films in the wide temperature range. Figures 2(a) and (b) show the results obtained from two representative films, with $x = 0.10$ and $0.17$. Both $M^S(T)$ and $R_{AH}^S(T)$ are measured by cooling from 400 K in a weak magnetic field $\mu_0 H_{FC} = 10$ mT. The first clear feature is that the AFM Néel transition takes place below 400 K in both films. It is noteworthy that the Néel transition in $Mn_3Sn$ is evidenced by a positive surge in magnetization. This ferromagnetic-like behavior arises from the combined effects of the formation of small WFMs and their unimpeded ordering along nonzero $\mu_0 H_{FC}$ at $T \cong T_N$. This simultaneously orders the octupole moments resulting in an equally abrupt emergence of $R_{AH}^S$.

The second important feature is the profound qualitative difference seen between these samples at low temperatures, starting below 170 K. We observe that whereas the $x = 0.10$ sample undergoes the $T_1$ transition from the inverse triangular antiferromagnetic to an incommensurate spin configuration[23] at about $T_1 = 150$ K, this transition is absent in the $x = 0.17$ film. Subsequently, the strong AHE signal characteristic for $Mn_3Sn$ is preserved down to the lowest temperatures in $Mn_{3.17}Sn_{0.83}$.

The sole weak temperature dependence of $R_{AH}^S$ for $Mn_{3.17}Sn_{0.83}$ down to 5 K [Fig. 2(b)] unequivocally demonstrates that the same negative polarity and substantial values of AHE remain unaltered until the lowest temperatures. To further substantiate this finding, we present in Figs. 3(a) and (b) the temperature evolution of AHE loops, $R_{AH}(H)$, for the same two samples considered in Figs. 2(a) and (b). The correlation with $T$-dependent data of Fig. 2 is indisputable. That is, for $Mn_{3.10}Sn_{0.90}$, the $T_1$ transition is observed, resulting in the disappearance of AHE at low temperatures, whereas for $Mn_{3.17}Sn_{0.83}$, it is absent, leading to a strong AHE in the whole temperature range down to 5 K.

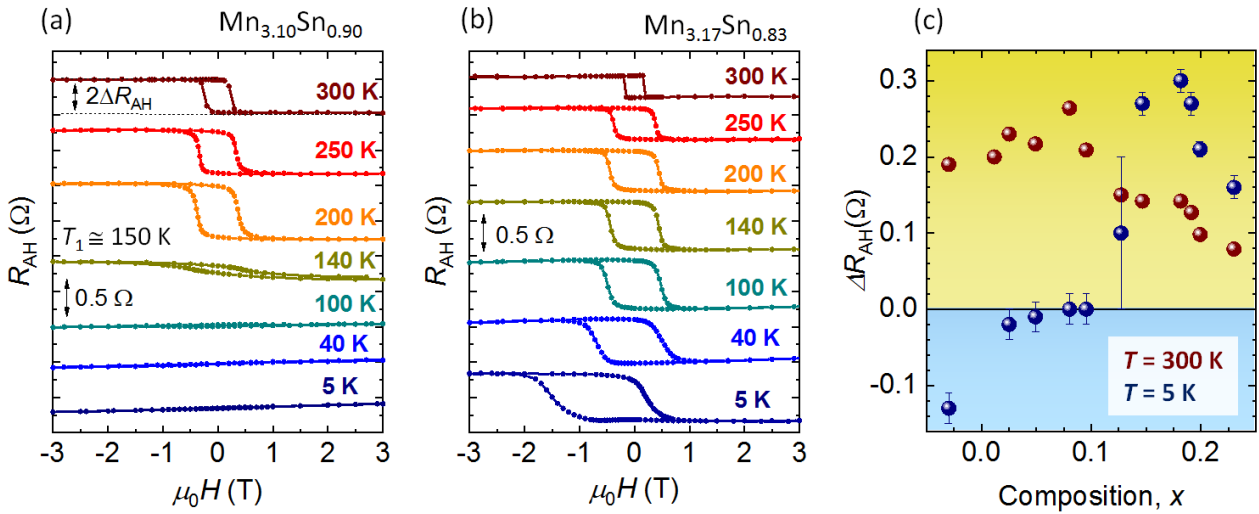

FIG. 3. (a,b) Temperature evolution of anomalous Hall resistance $R_{AH}$ in $Mn_{3+x}Sn_{1-x}$ films. The linear background of transverse resistance has been subtracted and the results are shifted vertically for clarity. The measurements are performed on warming after cooling at $\mu_0 H_{FC} = 1$ T. (c) Composition dependence of magnitudes of $\Delta R_{AH}$ [defined in (a)] established at 5 K (navy) and 300 K (brown). The sign of $\Delta R_{AH}$ represents the polarity of the AHE loop, such that the positive values on the graph correspond to negative polarity characteristic for its octupolar origin. The magnitude of the error bars is set by the residual curvature of $R_{AH}(H)$ at high magnetic fields.

The summary of the AHE results for the full $x$ range of studied samples, taken at 300 K and 5 K as the representative temperatures, is presented in Fig. 3(c). Firstly, we note that the magnetic octupole hosting $D0_{19}$-Mn$_3$Sn structure is realized for all compositions of Mn, including the below-stoichiometric one (see also supplementary information Fig. S3). As previously reported,[34] the strongest AHE at room temperature occurs in moderately doped films ($x \cong 0.1$). Temperature-dependent measurements reveal that the AHE diminishes at low temperatures for these concentrations. It is either quenched completely or acquires FM-like polarity; the latter phenomenon may be attributed to the presence of additional magnetic phases.[24,33,44] In contrast, all films with $x \geq 0.15$ exhibit a robust AHE response persisting down to at least 5 K, with the same negative polarity indicative of its octupolar origin. The Mn$_{3.12}$Sn$_{0.88}$ film appears to be a borderline case worthy of further investigation.

The results in Fig. 3(b) also point out one more intriguing phenomenon. The low-$T$ $R_{AH}(H)$, as well as $M(H)$ (not shown here), are biased in the opposite direction to the sign of $H$ in which the sample was initially cooled down, $\mu_0 H_{FC} = +1$ T, thereby resembling the classical exchange bias effect.[45] This feature is observed in all the samples in which the IT-AFM extends to the lowest temperatures. It is noteworthy that various biasing effects have been reported for bare Mn$_3$Sn even at room temperature.[46,47] In these cases, some kind of nanoscale phase coexistence was invoked to provide a firmly-biased pinning component. However, no evidence supporting such possibilities has been found in our films. A more in-depth investigation of the origin of this low-$T$ AHE shift is beyond the scope of this letter.

We summarize this part by noting that the accumulated body of evidence presented in Figs. 1-3 decisively indicates that a significant substitution of Sn by Mn in Mn$_3$Sn leads to the stabilization of the IT-AFM phase down to the lowest temperatures without any particular detrimental effects to the structural constitution of this compound.

## C. $T_N$ dependence on $x$

We now consider the variation of $T_N$ with $x$. The transition temperatures are determined from the positions of extrema in $dM^S(T)/dT$ and $dR_{AH}^S(T)/dT$ derivatives; the representative results for Mn$_{3.10}$Sn$_{0.90}$ are shown in Fig. 2(c). We underline here that both derivatives indicate the same characteristic temperatures, meaning that the measurements of either $M^S(T)$ or $R_{AH}^S(T)$ provide with equally accurate magnitudes of $T_N$ and $T_1$.

The dependence of $T_N$ on $x$ is shown in Fig. 4(a). Unlike in bulk crystals,[26] $T_N$ in our Mn$_{3+x}$Sn$_{1-x}$ thin films is sensitive to compositional changes. Close to the stoichiometric point, $T_N$ remains unchanged at around 390 K. However, as $x$ increases, $T_N$ drops significantly in a step-like manner, stabilizing at approximately 315 K for $x \geq 0.13$. This stabilization of $T_N$ corresponds to the solubility limit of Mn in Mn$_3$Sn, as evidenced by structural characterization [Fig. 1(f)]. The observed reduction of $T_N$ toward 300 K on increasing $x$ qualitatively explains the roll down of the AHE amplitudes at 300 K for $x \geq 0.12$ [Fig. 3(c)]. The corresponding reduction of $H_c$ is shown in Supplementary Information Fig. S3.

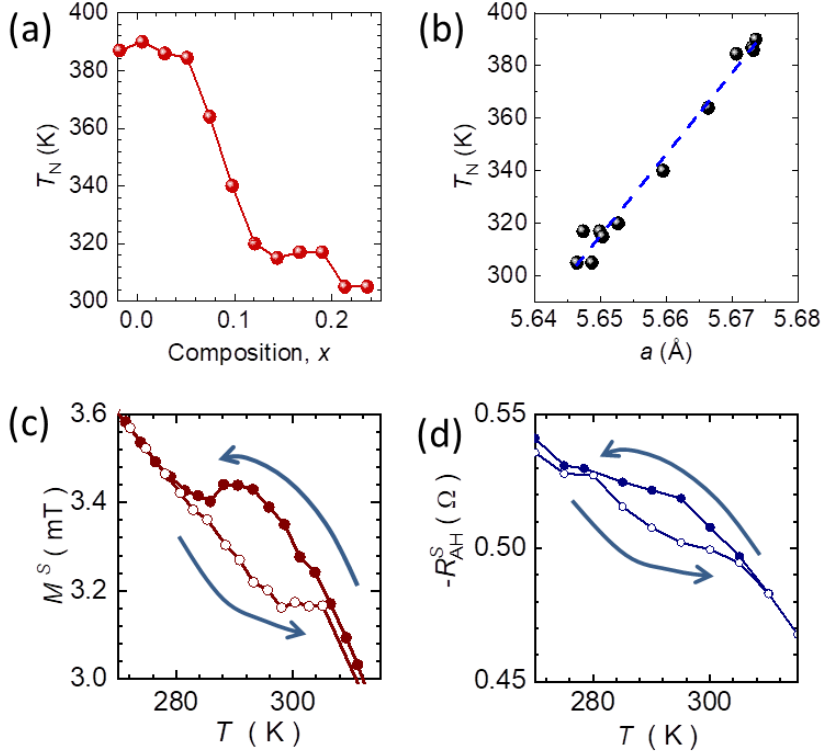

**FIG. 4.** The Mn$_{3+x}$Sn$_{1-x}$ Néel temperature, $T_N$, dependence on (a) composition, $x$, and (b) the $a$ lattice parameter. (c-d) Thermal hysteresis of spontaneous magnetization, $M^S(T)$, and anomalous Hall resistance, $R_{AH}^S(T)$, respectively, around $T_2$ anomaly for Mn$_{3.10}$Sn$_{0.90}$.

In general, the reduction of $T_N$ in thin films compared to the bulk value is often attributed to weakened magnetic exchange interactions, or surface and interface-related effects. In the case of Mn$_3$Sn, previous studies on bulk samples reported different values of $T_N$ between the surface and bulk regions.[22] In our study, the primary effects may be related to a strain-induced lattice distortion caused by both the reduced dimensionality and epitaxial strain, since $T_N$ in the studied films shows a clear correlation with $a$ lattice parameter, as depicted in Fig. 4(b). However, it must be acknowledged that a diffusion of atoms from adjacent layers within the stack cannot be ruled out.[48] The deposited layers are subjected to high-temperature annealing, which is essential to achieve the desired $m$-plane orientation of the films. Nevertheless, all layers are prepared under identical conditions in terms of stack design, thickness, and post-growth annealing conditions. As a result, any diffusion effects would systematically influence the absolute values of characteristic temperatures without altering the observed dependence on layer composition or *a*.

## D. Magnetic anomaly at $T_2$

The differentiation of the temperature-dependent data [Fig. 2(c)] clearly reveals the existence of a small magnetic anomaly between $T_1$ and $T_N$, labeled as $T_2$. This anomaly manifests as a distinct saddle point in either $M^S(T)$ or $R_{AH}^S(T)$, appearing roughly 55 K below $T_N$. Notably, this anomaly is ubiquitous, observed across all studied layers. Further investigation of $T_2$, as shown in Figs. 4(c-d), reveals thermal hysteresis in both $M$ and $R_{AH}$. The presence of thermal hysteresis suggests a first order transition at $T_2$.

Similar anomalies in magnetization near the transition temperature are often observed in soft ferromagnetic materials and are attributed to the Hopkinson effect,[49] which occurs as a ferromagnet is heated toward its Curie temperature $T_C$ due to increased domain wall mobility. However, in our study, the polarity of the AHE remains unchanged when crossing $T_2$. Therefore, rather than the domain pattern changes, it is more likely that only minor adjustments in the spin canting are involved. If so, these slightly different spin configurations must be separated by significant energy barriers to produce thermal hysteresis at such elevated temperatures.

We recognize that further experimental evidence is required to understand the origin of this anomaly and its potential role in Mn$_3$Sn-based devices. Nevertheless, the simultaneous appearance of such subtle features in both $M$ and the AHE underscores the close relationship between spin configuration details and the magnitude of the Berry curvature in Mn$_3$Sn.

## E. Mn$_{3+x}$Sn$_{1-x}$ magnetic phase diagram

Our letter concludes with Fig. 5, which depicts the magnetic phase diagram for epitaxial Mn$_{3+x}$Sn$_{1-x}$ films. We have established the position of the boundaries between three magnetic phases of Mn$_{3+x}$Sn$_{1-x}$ (paramagnetic, inverse triangular AFM, i.e. the commensurate with the lattice, and an incommensurate one) by determining the position of the characteristic points in $dM^S(T)/dT$ derivatives, as illustrated in Fig. 2(c). The extracted data show clear and smooth trends indicating in particular that in epitaxial Mn$_{3+x}$Sn$_{x-1}$ films, the $T_1$ transition from IT-AFM to the incommensurate phase can be continuously tuned down until its complete suppression by increasing the Mn-Sn composition above approximately $x = 0.13$. Furthermore, it is observed that the $T_2$ anomaly trails $T_N$, consistently remaining approximately 55 K below it. We postulate therefore, that the position of the $T_2$ anomaly can serve as a convenient indicator of $T_N$ in layers with $T_N$ exceeding 400 K, that is above the typical temperature limit for most popular experimental setups.

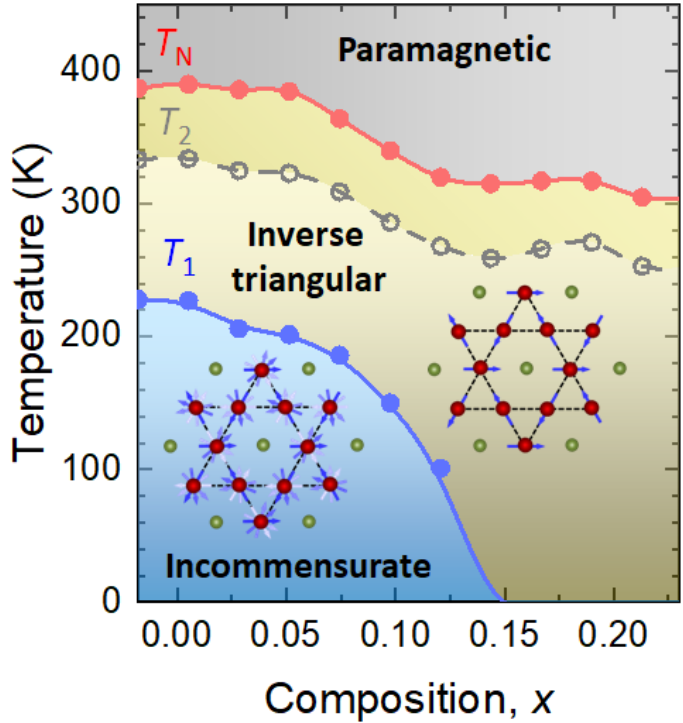

**FIG. 5.** The magnetic phase diagram of epitaxial of $Mn_{3+x}Sn_{1-x}$ thin films. The symbols marking the boundaries between magnetic phases are established at very weak external magnetic field, upon the positions of the characteristic points in the $dM^S(T)/dT$ derivatives, as illustrated in Fig. 2(c). The reddish bullets represent the Néel transition at $T_N$, while the blue ones mark the transition from the inverse triangular to an incommensurate spin arrangement at $T_1$. The open, gray symbols indicate the position of the magnetic anomaly, $T_2$, occurring consistently at 55 K below $T_N$ in all the layers under study. The cartoons illustrate the spin arrangements in the inverse triangular and incommensurate structures of $Mn_3Sn$. The red bullets represent Mn atoms, arrows their spins. The greenish bullets represent Sn atoms.

## IV. CONCLUSIONS

In summary, the study demonstrates that intrinsic alloying $Mn_3Sn$ with Mn enables the extension of the temperature range in which the unique topological and magnetic properties arising from inverse triangular antiferromagnetic (IT-AFM) structure can be harnessed. The results clearly demonstrate that, within the framework of sputtering deposition, the $T_1$ temperature – marking the transition from the IT-AFM structure to an incommensurate spin configuration - can be precisely tuned by adjusting the Mn-Sn composition, ultimately leading to its complete suppression. This, in turn, allows the strong anomalous Hall effect related to the magnetic octupole moment to persist down to the lowest temperatures. These findings open up new opportunities for applying the topological physics of $Mn_3Sn$ at low temperatures, or could even lead to entirely new research directions.

Furthermore, we demonstrate that the Néel temperatures in our stacks of *m*-plane $Mn_3Sn$ thin films are lower than the bulk values, and that they decrease in accordance with the changes of *a* lattice parameter as the Mn content is increased. Additionally, we report on the presence of a magnetic anomaly that appears consistently around 55 K below the Néel temperature, and on the thermal hysteresis associated with it. In this context, well-planned temperature-dependent studies are essential for gaining deeper insights into the actual spin and electronic structures of this material family. Such investigations should extend into the high-temperature range (above 300 K), as the reduction of the Néel temperature, observed here, could play a critical role in experiments where Joule heating is unavoidable. Additionally, the presence of thermal hysteresis - and associated energy barriers - warrants further exploration. The precise origin of this anomaly and its potential impact on device performance remain important questions for future research.

## ACKNOWLEDGEMENTS
The authors thank Y. Nakano for the measurement of inductively coupled plasma mass spectrometry (ICP-MS) and R. Nomura for Hall bars fabrication. This work was partly supported by TUMUG Support Program from Center for Diversity, Equity, and Inclusion, Tohoku University, JSPS KAKENHI (Grant Nos. 22K14558, 22KK0072, 24KJ0432, 24K22949, 24H02235 and 24H00039), JST-PRESTO (Grant No. JPMJPR24H6), JST-ASPIRE (Grant No. JPMJAP2322), MEXT Initiative to Establish Next-generation Novel Integrated Circuits Centers (X-NICS) (Grant No. JPJ011438), and RIEC Cooperative Research Projects.

# Supplementary information

## S. I. Surface and EDX studies

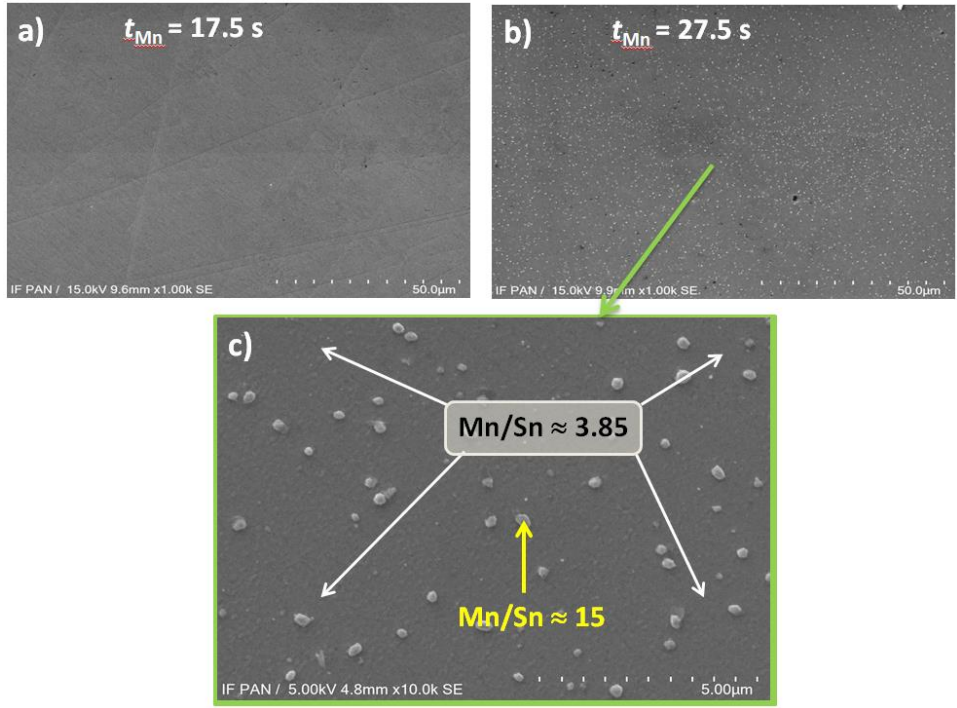

Fig. S1. (a-b) Scanning electron microscope images of $Mn_3Sn$ films with Mn single-target sputtering time, $t_{Mn}$, of 17.5 and 27.5 s. (c) A close-up of the film in (b) indicating locations and results of energy dispersive X-ray microanalysis.

Surface imaging and energy dispersive X-ray (EDX) microanalysis is performed using Hitachi SU-70 electron microscope. The most representative results are shown in Figure S1. Scanning electron microscopy of $Mn_{3+x}Sn_{1-x}$ thin films, sputtered with Mn single-target times of $t_{Mn}$ = 17.5 and 27.5 s (Figs. S1a-b), reveals the appearance of entities approximately 0.5 µm in diameter on the surface of the second film. These features are consistently observed in films with $t_{Mn}$ greater than 22.5 s, corresponding to $x > 0.13$, and their density increased with longer $t_{Mn}$. EDX analysis performed for these entities yields Mn/Sn ratio of about 15, meaning mostly Mn, whereas the Mn/Sn ratio established on the free surface is $3.85 \pm 0.04$ for this particular film, what correspond to $x = 0.18 \pm 0.03$. The last value aligns very well with $x = 0.16$ determined for this $t_{Mn}$ by inductively coupled plasma mass spectrometry on a reference set of samples, as can be read from Fig. 1(b) in the main text.

## S. II. Detrimental role of the magnetic response of MgO substrates on thin film magnetometry.

Ignoring the actual shape of the magnetic response of common commercial substrates makes the traditional approach to thin film magnetometry prone to significant qualitative errors [1,2]. When thin films exert a weak magnetic moment, such as when they are magnetically diluted [3] or antiferromagnetic, problems inevitably arise. The use of oxide compounds as substrates for deposition makes magnetometry particularly challenging due to the presence of two strong non-diamagnetic contributions, a paramagnetic-like and a ferromagnetic-like [4], which considerably distort the native diamagnetism of the host lattice of the substrate material.

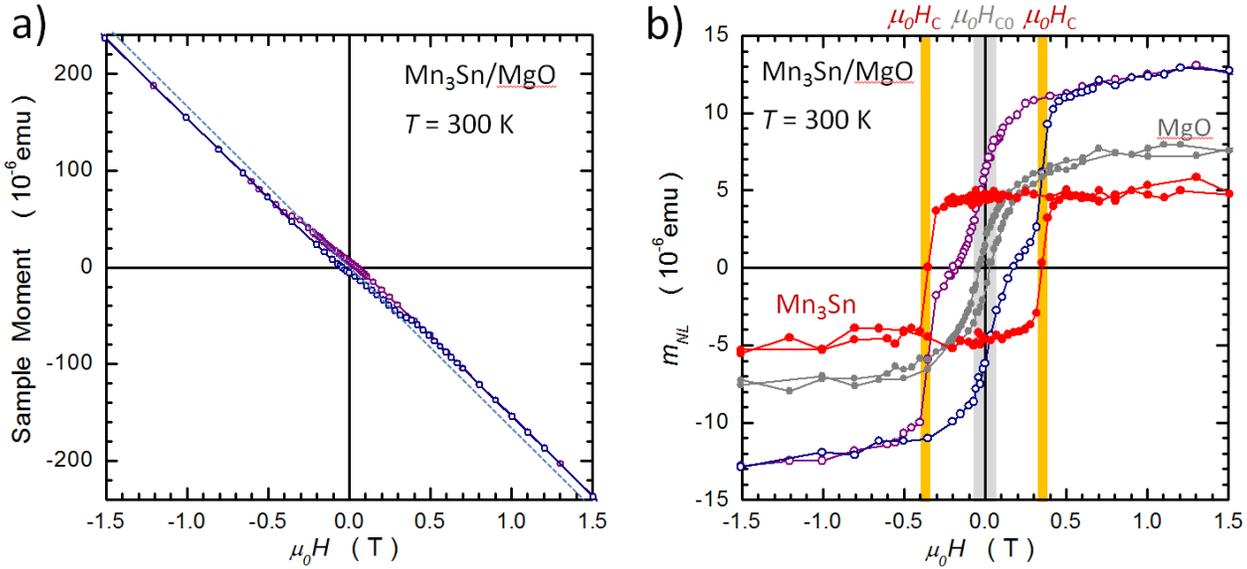

Fig. S2. (a) Magnetic field, $\mu_0 H$, dependent raw magnitudes of magnetic moments of the test sample: (open symbols) 50 nm thick Mn$_3$Sn film deposited on 5×5×0.5 mm$^3$ MgO substrate of a mass $\nu \cong 43$ mg). The dashed line marks the expected *diamagnetic* response of the MgO substrate: $\nu \chi_{\mathrm{MgO}} H$, where $\chi_{\mathrm{MgO}}$ is the diamagnetic susceptibility of MgO. b) The open and full grey symbols mark the nonlinear parts of the magnetic moments, $m_{NL}$, for the results presented in panel (a) and for the similar piece of MgO substrate, respectively, that is after subtraction of the linear contribution $\nu \chi_{\mathrm{MgO}} H$. The red bullets mark the difference between the two sets of results. Measurements are performed with $H$ applied perpendicularly to the sample plane, that is, in the case of Mn$_3$Sn being parallel to the kagome planes and the octupole moments. All the data are shown in experimental units to straightforwardly indicate the real scale of the magnetic distortions brought about by the unexpected strong nonlinear in $H$ contribution present in commercial MgO substrates.

Fig. S2 exemplifies the significant and detrimental influence of the MgO substrates in determining the true magnetic moment of Mn$_3$Sn films. Fig. S2(a) shows the raw magnetometry data (dark open symbols) for a 50 nm thick test Mn$_3$Sn film deposited in the same structure as in the main study, as shown in Fig. 1(a) in the main text. The magnetic response is clearly dominated by that of the substrate. This is an expected outcome, of course. The problems mount when it is assumed that the MgO substrate contributes only the ideal diamagnetic response, that is proportional to the applied magnetic field: $\nu\chi_{MgO}H$, where $\chi_{MgO}$ is the diamagnetic susceptibility of MgO and $\nu$ is the mass of the specimen. Such a contribution is indicated by the dashed straight line in Fig. S2(a). In Fig. S2(b), using the same symbols as in Fig. S2(a), we plot the difference between the experimental points form Fig. S2(a) and $\nu\chi_{MgO}H$ straight line. A complex hysteresis curve is obtained below 0.5 T, with double switching events at $\mu_0 H_C = \pm 0.35$ T (indicated by orange bars) and $\mu_0 H_{C0} \cong 0$ T indicated by the light gray bar.

It turns out that the second switching event around $\mu_0 H_{C0} \cong 0$ T and the subsequent curvature of the magnetic response is caused solely by the non-diamagnetic component introduced the MgO substrate alone. It is depicted using full gray symbols in Fig. S2(b). It has been obtained by subtracting the same straight line dependency, $\nu\chi_{MgO}H$, from the raw magnetometry data of the base MgO substrate (as obtained from a vendor) measured in the same magnetic configuration to that of the test sample. We can clearly recognize that both the switching at weak magnetic fields (with a coercivity of about 0.03 T) and a rounded magnetization rise towards saturation at about 1 T are present in the MgO substrate piece.

The resulting magnetization curve of the $Mn_3Sn$ layer is marked by the red bullets in Fig. S2(b). This curve is obtained by subtracting the measured MgO-substrate magnetization curve from that of the whole sample. As shown in Fig. 1(g) of the main text, the dependence is qualitatively similar to that of the $Mn_{3.07}Sn_{0.93}$ film, with a single switching event at $H_c$ and a flat (linear in $H$) magnetic response away from reversal events. Notably, the first switching event around $\mu_0 H_{C0} \cong 0$ T has been completely eradicated from the final results.

Returning to the MgO, the example shown here strongly emphasize the fact that these substrates bring an additional sizable non-diamagnetic response characterized by soft magnetic hystereses that slowly saturate at about 1 T at a very high level of about $10^{-5}$ emu per typical 5×5 $mm^2$ sample piece, and thus may dwarf the sought-after magnetic contribution from the thin film.

Prospective investigators must at least follow the experimental pattern sketched above or adopt the full experimental method of *in situ* compensation of unwanted substrate signals [5] to achieve the best results. We underline the fact that the *in situ* compensation technique was already proven very effective in *quantitative* studies of other AFM thin films [6–9].

Finally, we emphasize that the features presented above are general and not specific exclusively to the MgO substrates which are used in the current study, or the vendor from which they were acquired.

# S. III. Composition dependence of the coercive field, $H_c$ at 300 K.

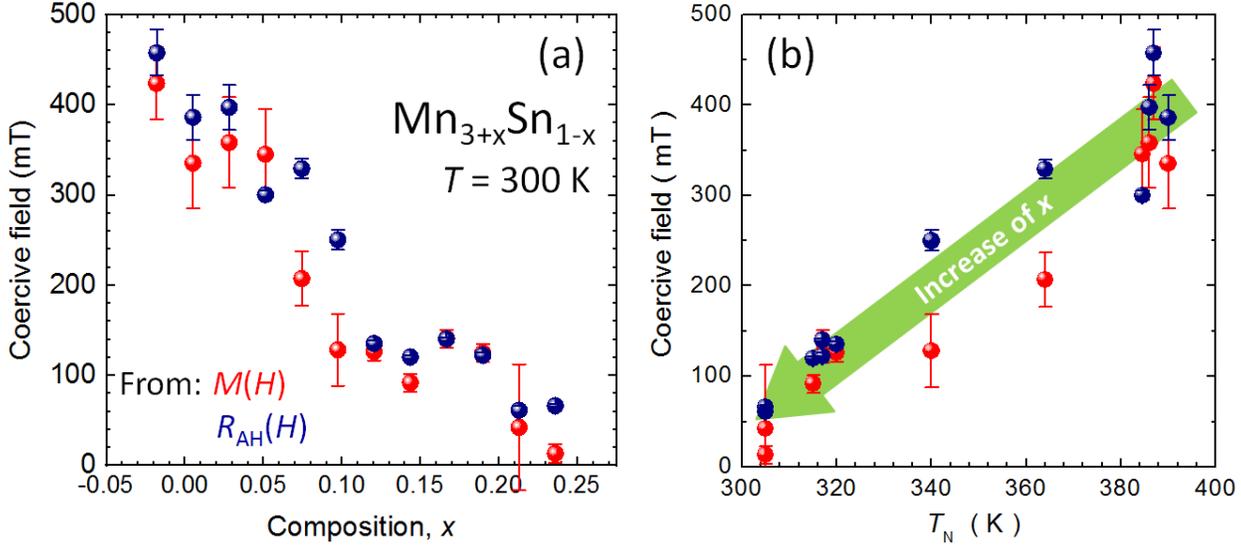

Figure S3. (a) Composition $x$ dependence of the coercive field, $H_c$, established from magnetization (red bullets) and anomalous Hall effect (navy bullets) magnetic loops at $T = 300$ K. (b) The same, but as a function of the Néel temperature of the given film taken from Fig. 4(a) of the main text. The background arrow indicated the direction of the increasing $x$.

There are two main effect which stem out from the data depicted in Fig. S3(a):

1) That the anomalous Hall effect of the inverted polarity corresponding to the octupolar origin is seen in $Mn_{3+x}Sn_{1-x}$ layers at room temperature for all the concentration studies, that is from the below stoichiometric $x = -0.02$ up to about $x = 0.23$.

2) There exists a nearly perfect correspondence of the values of the coercive field, $H_c$, as determined from magnetic $M(H)$ curves and anomalous Hall effect $R_{AH}(H)$ curves extracted from transverse magnetoresistance measurements. This highlights the full correspondence of magnetometric and magnetoresistance methods in determination of the magnetic constitution of $Mn_3Sn$.

3) The roll off the magnitudes of the coercive fields established at 300 K for increasing $x$ in the films aligns with the reduction of the Néel temperature, $T_N$, towards 300 K observed for these layers [shown in Fig. 4(a) in the main text]. The existence of a clear $H_c \Leftrightarrow T_N$ correlation is documented in Fig. S3(b).

# S IV. Supplementary information references